\documentstyle[12pt,epsf]{article}

\newcommand{\nc}{\newcommand}
\nc{\fdiag}{0}
\nc{\bg}{B. Grz${{{\rm a}_{}}_{}}_{\hskip -0.18cm\varsigma}$dkowski}
\nc{\lsp}{\;\;\;\;\;\;\;\;}
\nc{\beq}{\begin{equation}}   \nc{\eeq}{\end{equation}}
\nc{\bea}{\begin{eqnarray}}   \nc{\eea}{\end{eqnarray}}
\nc{\baa}{\begin{array}}      \nc{\eaa}{\end{array}}
\nc{\bit}{\begin{itemize}}    \nc{\eit}{\end{itemize}}
\nc{\ben}{\begin{enumerate}}  \nc{\een}{\end{enumerate}}
\nc{\bce}{\begin{center}}     \nc{\ece}{\end{center}}
\nc{\non}{\nonumber}
\nc{\lumun}{\;{\hbox {pb}^{-1}}{\hbox {yr}^{-1}}}
\nc{\hc}{\hbox {h.c.}}
\nc{\re}{\hbox {Re}}
\nc{\im}{\hbox {Im}}
\nc{\etal}{\hbox{et al.}}
\nc{\prdj}[3]{{\it Phys.\ Rev.}\ {{\bf D{#1}} (#2), #3}}
\nc{\prlj}[3]{{\it Phys.\ Rev.\ Lett.}\ {{\bf {#1}} (#2), #3}}
\nc{\plbj}[3]{{\it Phys.\ Lett.}\ {{\bf B{#1}} (#2), #3}}
\nc{\npbj}[3]{{\it Nucl.\ Phys.}\ {{\bf B{#1}} (#2), #3}}
\nc{\ptpj}[3]{{\it Prog.\ Theor.\ Phys.}\ {{\bf {#1}} (#2), #3}}
\nc{\zfpj}[3]{{\it Z.\ Phys.}\ {{\bf C{#1}} (#2), #3}}
\nc{\mplaj}[3]{{\it Mod.\ Phys.\ Lett.}\ {{\bf A{#1}} (#2), #3}}
\nc{\rmpj}[3]{{\it Rev.\ Mod.\ Phys.}\ {{\bf {#1}} (#2), #3}}
\nc{\ijmpaj}[3]{{\it Int.\ J.\ of\ Mod.\ Phys.}\ {{\bf A{#1}} (#2), #3}}
\nc{\ra} {\rightarrow}
\nc{\cw}{\cos\theta_W}        \nc{\sw}{\sin\theta_W}
\nc{\ttbar}{t\bar{t}}
\nc{\bbbar}{b\bar{b}}
\nc{\tanb} {\tan \beta}
\nc{\twbdec} {t\rightarrow W^+ b}
\nc{\tbwbdec} {\bar{t} \rightarrow W^- \bar{b}}
\nc{\hprod} {e^+e^- \ra Z^\ast \ra H Z}
\nc{\epem} {e^+e^-}
\nc{\wpwm} {W^+W^-}
\nc{\tbar} {\bar{t}}
\nc{\bbar} {\bar{b}}
\nc{\wpp} {W^+}
\nc{\mt}{m_t}
\nc{\mts}{m_t^2}
\nc{\mw} {m_W}
\nc{\mws} {m_W^2}
\nc{\mz} {m_Z}
\nc{\mzs} {m_Z^2}
\nc{\mh} {m_H}
\nc{\mhs} {m_H^2}
\nc{\ma} {m_A}
\nc{\mas} {m_A^2}
\nc{\hdec}{H \ra t\bar{t}}
\nc{\ttbardec}{\ttbar \ra W^+W^-\bbbar}
\nc{\po}{\Phi_1}
\nc{\pod}{\Phi_1^\dagger}
\nc{\pht}{\Phi_2}
\nc{\phtd}{\Phi_2^\dagger}
\nc{\phtt}{{\tilde{\Phi}}_2}
\nc{\popo}{\po^\dagger\po}
\nc{\phtpt}{\pht^\dagger\pht}
\nc{\popt}{\po^\dagger\pht}
\nc{\phtpo}{\pht^\dagger\po}
\nc{\sq}{\sqrt{2}}
\nc{\nsd} {N_{SD}}
\nc{\ntt} {N_{tt}}

\def\lammsbar{\Lambda_{\overline{\rm MS}}}
\def\lsim{\mathrel{\raise.3ex\hbox{$<$\kern-.75em\lower1ex\hbox{$\sim$}}}}
\def\gsim{\mathrel{\raise.3ex\hbox{$>$\kern-.75em\lower1ex\hbox{$\sim$}}}}
\def\anti{\overline}

\def\gev{\,{\rm GeV}}

\def\wt{\widetilde}

\def\anti{\overline}

\def\h{h}
\def\a{a}
\def\mh{m_{\h}}
\def\ma{m_{\a}}

\def\etal{{\it et al.}}
\def\epem{e^+e^-}

\def\lsim{\mathrel{\raise.3ex\hbox{$<$\kern-.75em\lower1ex\hbox{$\sim$}}}}
\def\gsim{\mathrel{\raise.3ex\hbox{$>$\kern-.75em\lower1ex\hbox{$\sim$}}}}
\def\@versim#1#2{\vcenter{\offinterlineskip
        \ialign{$\m@th#1\hfil##\hfil$\crcr#2\crcr\sim\crcr } }}

\def\ie{{\it i.e.}}

\def\nsd{N_{SD}}
\def\anti{\overline}

\def\gev{\,{\rm GeV}}

\def\wt{\widetilde}

\def\sq{\wt q}

\def\tanb{\tan\beta}

\def\mt{m_t}

\def\mc{m_c}
\def\mz{m_Z}
\def\mw{m_W}

\def\h{h}
\def\mh{m_{\h}}

\begin{document}
%
\font\fortssbx=cmssbx10 scaled \magstep2
\medskip
$\vcenter{
\hbox{\fortssbx University of California - Davis}
}$
\hfill
$\vcenter{
\hbox{\bf UCD-97-14} 
\hbox{\bf LBNL-40399}
\hbox{June, 1997}
}$
\vspace*{1cm}
\begin{center}
{\large{\bf Intrinsic Charm at High-$Q^2$ and HERA Data}}\\
\rm
\vspace*{1cm}
{\bf  J.F. Gunion$^a$ and R. Vogt$^{a,b}$}\\
\vspace*{1cm}
{$^a$ \em Department of Physics, 
University of California, Davis, CA, USA }\\
{$^b$ \em Lawrence Berkeley National Laboratory, Berkeley, CA, USA}
\end{center}
\begin{abstract}
We compute the predictions of 
intrinsic charm for deep-inelastic scattering at high-$Q^2$ and compare
to HERA data. With the inclusion of constraints
from low-energy data, enhancements beyond the predictions of standard 
structure functions are very modest, but peaking in
the leptoquark mass variable is present near
200 GeV. Ultimately, the ability of HERA to probe the intrinsic
charm hypothesis could be very substantial.
\end{abstract}
\vspace{5mm}

\setcounter{page}{0}
\thispagestyle{empty}
 
HERA deep-inelastic scattering
data with significant statistics at high $Q^2$ and high $x$
is now becoming available~\cite{h1,zeus}.  
The cross section exhibits an excess beyond the predictions based
on standard QCD distribution functions. 
However, it is possible that unexpected contributions to
the quark distribution functions might explain at least part of this 
excess~\cite{tung}.
In this Letter, we pursue the implications for the HERA data
of an intrinsic charm (IC) component of the proton~\cite{icharm}.
We find that intrinsic charm
naturally predicts a peak in the leptoquark mass distribution in
the vicinity of 200 GeV, but that
existing constraints on intrinsic charm limit the size of the effect.
As a function of leptoquark mass, the enhancement from intrinsic charm
relative to standard model (SM) predictions is at most of order 15\%
in the $e^+p$ neutral current (NC) case but can be as large as 75\%
in $e^+p$ charged current (CC) scattering.
Expectations for $e^-p$ NC and CC scattering are also given.

The possibility of an intrinsic charm component in the proton bound
state has substantial theoretical and phenomenological motivation.
However, a conclusive observation or severe limit has remained elusive.
In the intrinsic charm model, there is a $|uudc\bar c\rangle$
component of the proton in which the $c$ and $\bar c$ have become
completely integrated members of the bound state, travelling coherently
with the light quarks.
In the minimal model, the $|uudc\bar c\rangle$ wave function is
taken to be proportional to the inverse of the light-cone energy
difference $m_p^2-\sum (m_i^2/x_i)\propto x_cx_{\bar
c}/(x_c+x_{\bar c})$\, where the $x_i$ are the light-cone momentum fractions
and the light quark and proton masses are neglected compared to
the charm quark mass. After squaring, 
the intrinsic charm Fock state probability distribution takes the form
\beq
{dP_{ic}\over dx_udx_u^\prime dx_ddx_cdx_{\bar c}}=
N_5 {x_c^n x_{\bar c}^n\over
(x_c+x_{\bar c})^n}\delta(1-\sum_i x_i)\,,
\label{pform}
\eeq
with $n=2$.
However, more extreme dependence on the light-cone energy difference is entirely
possible and, indeed, very natural, given that the strong interactions
will tend to bind the intrinsic charm quarks to the light quarks so that all
move with essentially the same velocity.
To exemplify this situation, we will also consider $n=8$ in Eq.~(\ref{pform}).

The inclusive charm quark distribution, $c(x)$, is obtained 
by integrating Eq.~(\ref{pform}) over all the $x_i$ except $x\equiv x_c$.
For $n=2$ and $n=8$, one finds
\beq
c(x)={1\over 2}N_5 x^2\left[{1\over 3}(1-x)(1+10x+x^2)+2x(1+x)\ln x\right]
~~~(n=2)\,,
\label{cform}\\
\eeq
\bea
c(x)&=&{1\over 210}N_5x^8\Biggl[35+1155x-1575x^2-11375x^3-2450x^4+490x^5
-96x^6+14x^7-x^8\nonumber\\
&&\phantom{{1\over 210}N_5} 
+x\left(1443+7161x+5201x^2+\left\{840+5880x+5880x^2\right\}\ln x\right)
\Biggr]~~~(n=8)\,,
\eea
respectively.
We adopt a normalization $N_5$ such that the intrinsic charm Fock state
component of the proton has 1\% probability, implying
$N_5=36$ ($N_5=2888028$) for $n=2$ ($n=8$).
The anti-charm distribution, $\bar c(x)$, is identical to $c(x)$.
Thus, at leading order, the net contribution from intrinsic $c$ and $\bar c$ 
to the deep-inelastic electromagnetic 
structure function is $F_2^{c~{\gamma^\star}}(x)=8xc(x)/9$.
Next-to-leading order QCD evolution corrections to the IC
contribution to $F_2^{c~{\gamma^\star}}(x)$ are 
computed following the formalism outlined in Ref.~\cite{hsv}. (See also
Ref.~\cite{hm}.) They
are significant but evolve very slowly with $Q^2$.

\begin{figure}[h]
\leavevmode
\epsfxsize=6.5in
\centerline{\epsffile{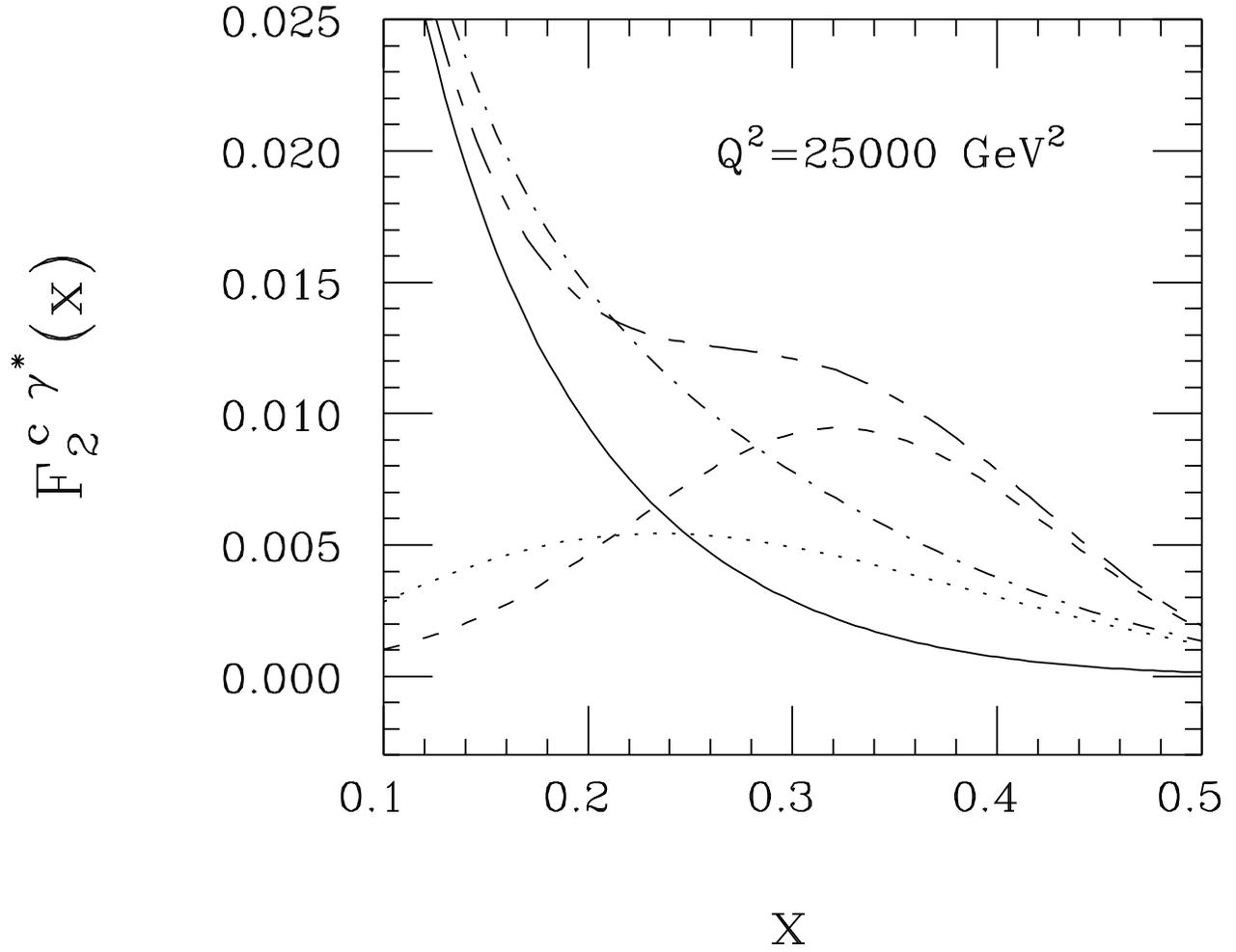}}
\bigskip
\caption{Predictions for $F_2^{c~{\gamma^\star}}(x)$ at $Q^2=25000\gev^2$:
solid --- SM next-to-leading order
prediction using MRS96(R2)~\protect\cite{mrs96} distribution functions only;
dots --- the $n=2$ IC component alone;
dot-dashed --- SM plus $n=2$ IC; 
dashes --- $n=8$ IC only;
dash-long-dash --- SM plus $n=8$ IC.}
\label{f2c}
\end{figure}

In Fig.~\ref{f2c}, we illustrate results 
for the electromagnetic structure function, $F_2^{c~{\gamma^\star}}(x)$ 
at $Q^2=25000\gev^2$. First, we give
the standard next-to-leading order
prediction for $F_2^{c~{\gamma^\star}}$ obtained by employing the recent
MRS96(R2)\cite{mrs96} distribution functions computed via
perturbative QCD evolution, without inclusion of the IC component.
Also shown is the $n=2$ IC component alone and the SM plus 
$n=2$ IC sum.\footnote{In our QCD evolution computations
for the intrinsic charm component, we
employ $\lammsbar=0.344\gev$, as appropriate for MRS96(R2), and $\mc=1.5\gev$.
We use scale $Q$ in evaluating distribution functions and QCD-evolution
corrections to $F_2^{c~{\gamma^\star}}(x)$.}
A very modest enhancement in the vicinity of $x\sim 0.2$ is observed.
Results for the $n=8$ model for the 
intrinsic charm component and corresponding total 
$F_2^{c~{\gamma^\star}}(x)$ are also
shown. A much more substantial enhancement is observed,
peaking in the vicinity of $x\sim 0.3$. 

\begin{figure}[h]
\leavevmode
\epsfxsize=6.5in
\centerline{\epsffile{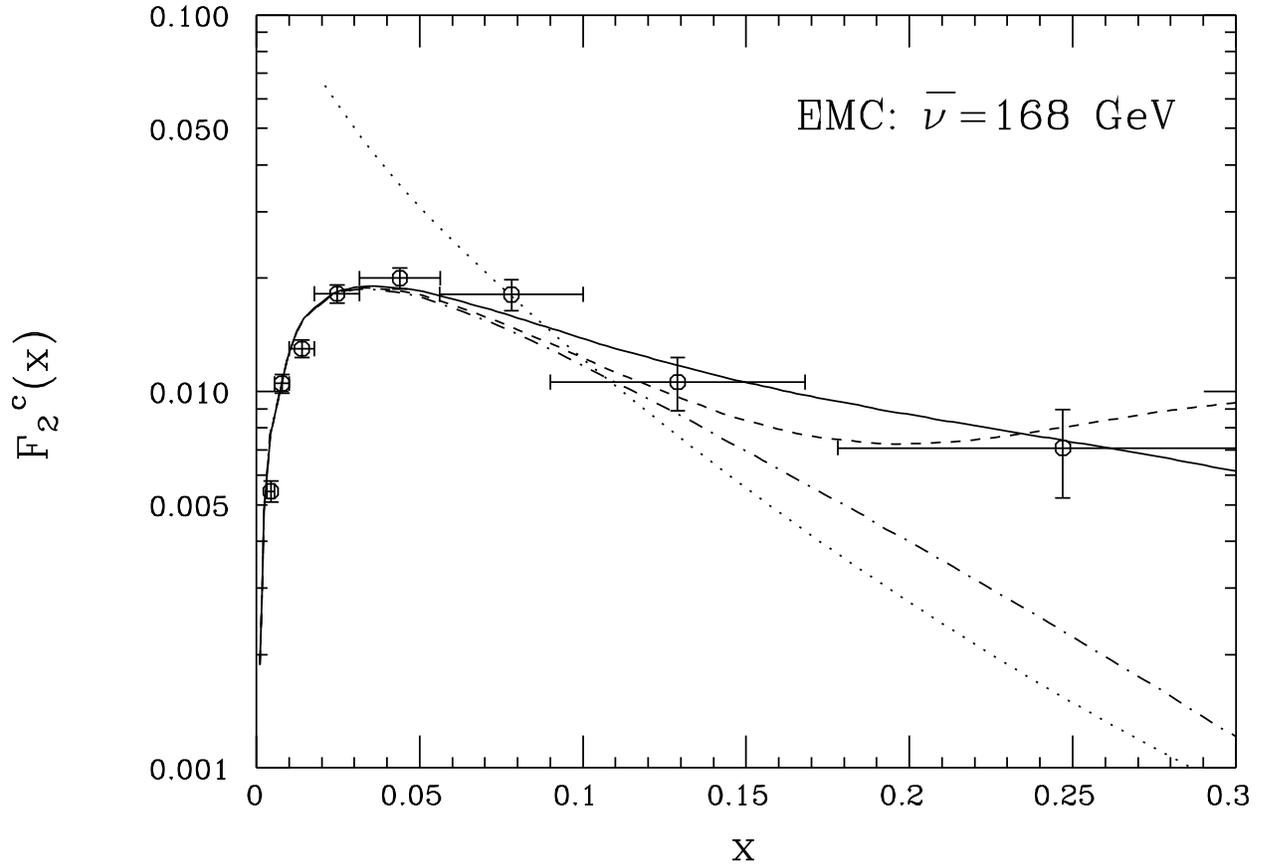}}
\bigskip
\caption{EMC data for $F_2^c(x)$ at $\overline\nu=168\gev$
compared to: (i) dotdash --- the extrinsic charm prediction of 
Ref.~\protect\cite{hsv}; (ii) dots --- the CTEQ3 perturbative 
prediction; (iii) solid --- EC+IC prediction for $n=2$; (iv) dashes ---
EC+IC prediction for $n=8$.}
\label{emccomp}
\end{figure}

The low-energy direct EMC measurement\cite{emc} of $F_2^c(x)$ provides
some evidence for intrinsic charm.  In Fig.~\ref{emccomp}, the EMC data for 
$\overline\nu=168\gev$ is compared
to several predictions for $F_2^c$.\footnote{At the $Q^2$
values probed by EMC, the $Z^\star$ contribution to $F_2^c$ is
negligible compared to the $\gamma^\star$ contribution.}
The first is the
perturbative ``extrinsic'' charm prediction of Ref.~\cite{hsv}.
In this approach, charm production in deep-inelastic scattering is computed 
to next-to-leading order using all relevant partonic-level
subprocesses contributing to
$\gamma^* p \to c X$. (In our comparison to this data, 
we employ the $\lammsbar=0.239\gev$
CTEQ3~\cite{cteq3} distribution set and use scale $\mu_0^2=Q^2+M_{c\anti c}^2$
in all distribution function and evolution evaluations~\cite{hsv}.) The most
important aspect of the subprocess approach is that 
charm-quark mass effects near threshold are correctly incorporated. This
is not true of the 
perturbative charm distributions contained in the CTEQ3 distribution set
itself; these are obtained by effectively assuming that the charm
quarks are massless. That correct inclusion of mass effects
is critical in the EMC energy range is shown
by the curve in Fig.~\ref{emccomp} corresponding to the CTEQ3 charm distribution
function. The other curves in Fig.~\ref{emccomp} illustrate
the result of adding intrinsic charm
components with $n=2$ and $n=8$ to the extrinsic charm prediction.\footnote{
In the $n=2$ IC case, the best two-component EC+IC fit is obtained for
$(1.27\pm0.06) F_2^{\rm EC}+ (0.92\pm0.53) F_2^{\rm IC}$. The EC coefficient,
1.27, is not inconsistent
with the expected $K$ factor from higher order corrections, while 
the IC coefficient is consistent with a $1\%$ IC probability. For the
$n=8$ curves, we have employed the same coefficients.}
In agreement with Ref.~\cite{hsv}, the $n=2$
addition to the extrinsic charm distribution fits the data quite nicely.
The $n=8$ model provides equally good agreement with the data.
Even larger values of $n$ would still be consistent with the EMC data,
but substantially larger probability for the IC wave function component
would not.

\begin{figure}[h]
\leavevmode
\epsfxsize=6.5in
\centerline{\epsffile{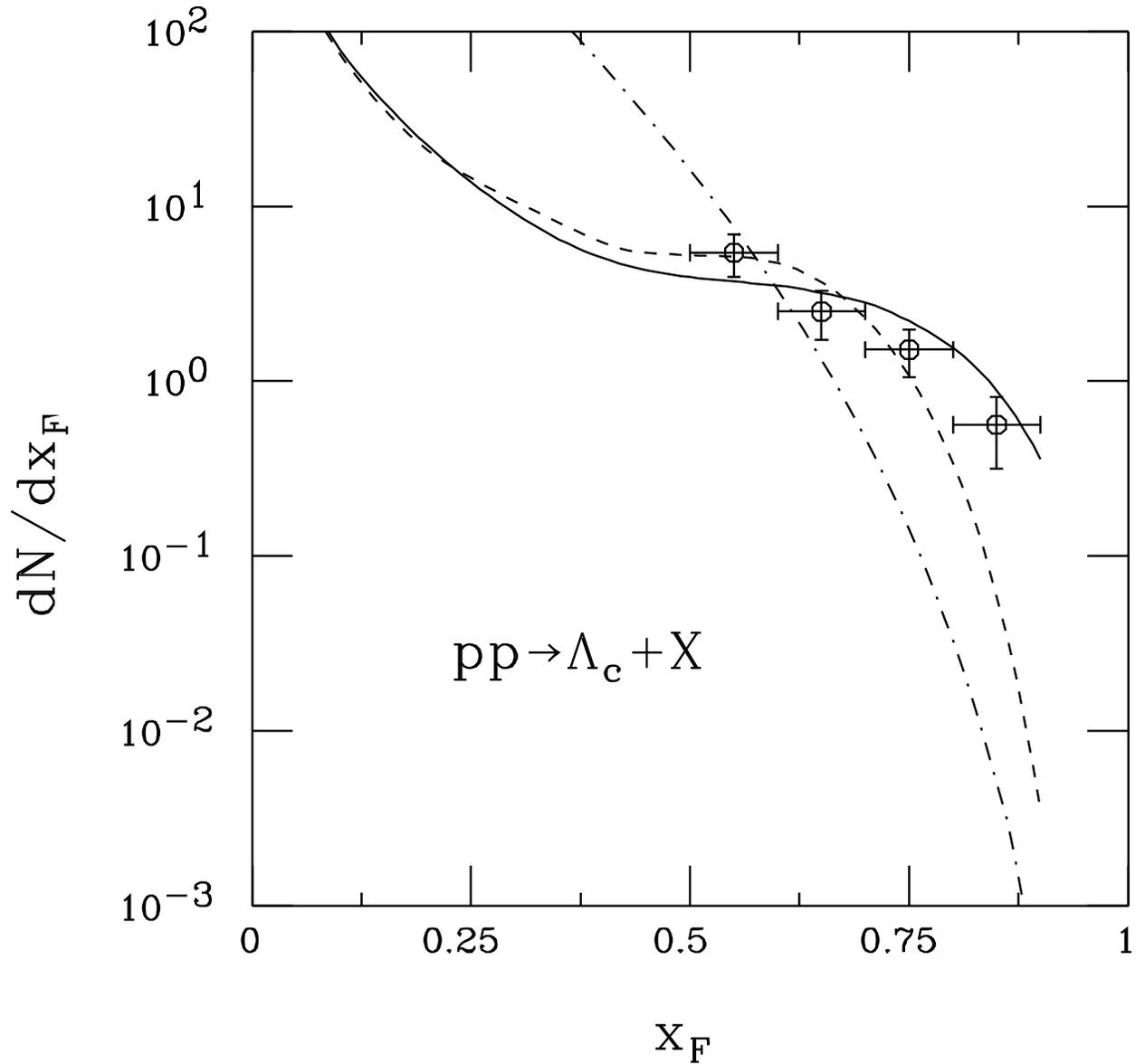}}
\bigskip
\caption{The $x_F$ distribution for $pp\to \Lambda_c+X$. Data from
Ref.~\protect\cite{lambdacisr} is compared to:
(i) dotdash --- $gg+q\anti q\to c\anti c$ fusion followed by $c\to\Lambda_c$;
(ii) solid --- fusion plus
$n=2$ intrinsic charm contributions;
(iii) dashes --- fusion plus $n=8$ IC contributions. A
$1\%$ probability for the IC component of the proton wave function is used
to fix the IC cross section.
In all three cases, the overall normalization is fixed by $\sigma(x_F\geq0.5)$.}
\label{lambdac}
\end{figure}

ISR data on $pp\to \Lambda_c+X$~\cite{lambdacisr} 
provides additional evidence for and constraints upon intrinsic 
charm~\cite{lambdacref1,lambdacref2}. The production of $\Lambda_c$ baryons
can be viewed as arising from three sources. 
First, there is $gg+q\anti q$ fusion production of a $c\anti c$ pair, 
followed by $c\to \Lambda_c$. Second, in inelastic $pp$ collisions
the $c$ quark in the IC proton wave function component\footnote{It is
assumed that there 
is no extrinsic or perturbative $c$ quark component to the proton wave function
at the low momentum transfers involved.} can fragment to a $\Lambda_c$.
Finally, also in inelastic collisions,
the $c$ quark in the IC component of the proton wave function can
coalesce with a $u$ and $d$ quark to produce a $\Lambda_c$; the
$x_F$ distribution of the $\Lambda_c$ produced in this manner
is obtained by integrating over all the parton momentum
fractions in Eq.~(\ref{pform}) with the constraint $x_F=x_u+x_d+x_c$.
Following the procedures of Refs.~\cite{lambdacref1,lambdacref2}, 
it is found that the shape of $dN/dx_F$ at large $x_F\geq 0.5$
is dominated by the third, \ie\ coalescence, term.
This is illustrated in Fig.~\ref{lambdac}, where $dN/dx_F$ is plotted
for (i) fusion, (ii) fusion plus $n=2$ IC and (iii) fusion plus $n=8$ IC.
In each case, the prediction is
normalized to $\sigma(x_F\geq 0.5)$.\footnote{Comparisons based
on absolute normalizations are less reliable.}
We observe that the $dN/dx_F$ data requires the presence
of an intrinsic charm contribution, and that the $n=2$ shape is preferred over
the $n=8$ shape. It is for this reason that we have chosen not to
consider still higher $n$ values in Eq.~(\ref{pform}) for the IC model,
despite the fact that higher $n$ values would yield larger effects at HERA.

\begin{figure}[h]
\leavevmode
\epsfxsize=6.5in
\centerline{\epsffile{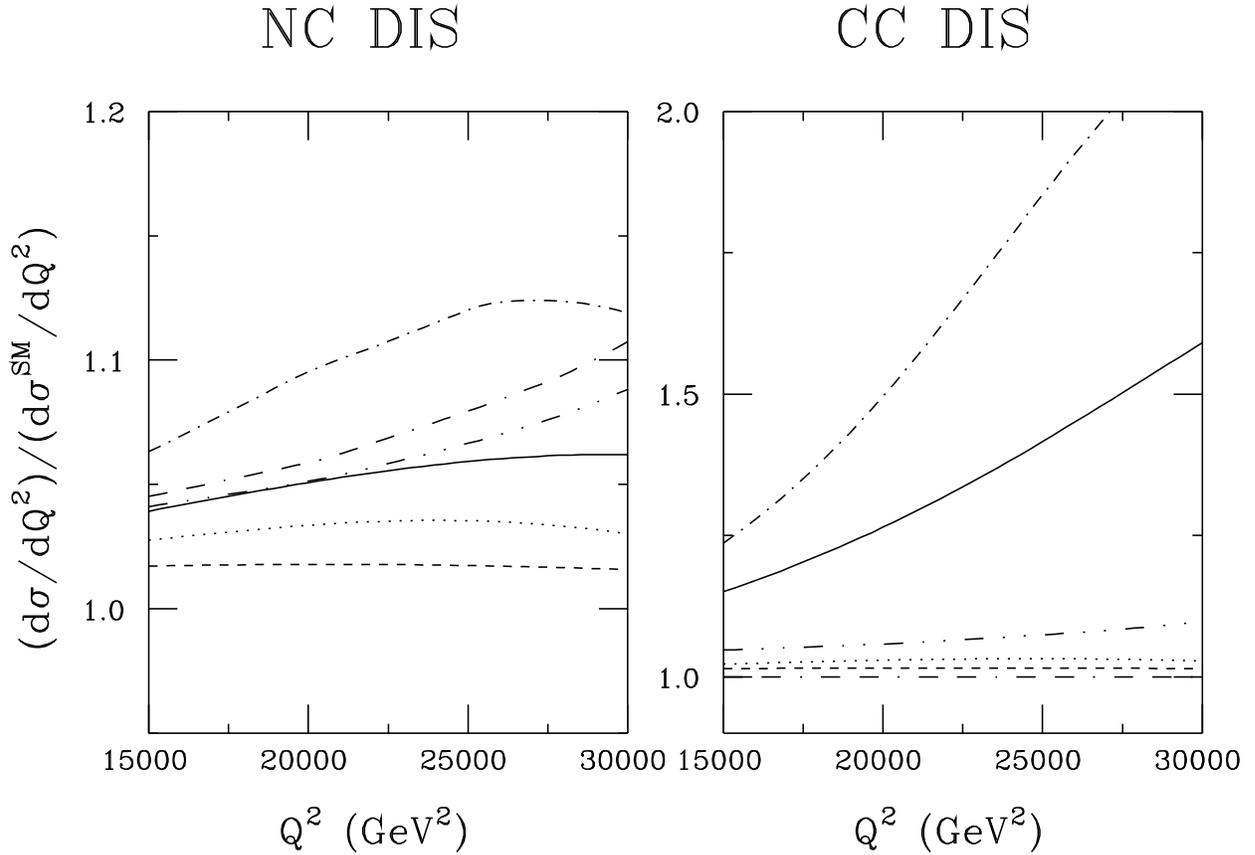}}
\bigskip
\caption{Predictions of various models for $[d\sigma/dQ^2]/[d\sigma^{SM}/dQ^2]$ 
at HERA center-of-mass energy
after integrating over $0.1\leq y\leq 0.9$.
Results are shown for $e^{\pm}p$ scattering and both neutral
current (NC) and charged current (CC) scattering.
Curve legend: solid --- $e^+p$ scattering and $n=2$ IC;
dotdash --- $e^+p$, $n=8$ IC; dashes --- $e^-p$, $n=2$ IC; dots $e^-p$,
$n=8$ IC; dash-dash-dot --- $e^+p$, $\delta u(x)=0.02(1-x)^{0.1}$; 
dash-dot-dot --- $e^-p$, $\delta u(x)=0.02(1-x)^{0.1}$.}
\label{rvsqsq}
\end{figure}

\begin{figure}[h]
\leavevmode
\epsfxsize=6.5in
\centerline{\epsffile{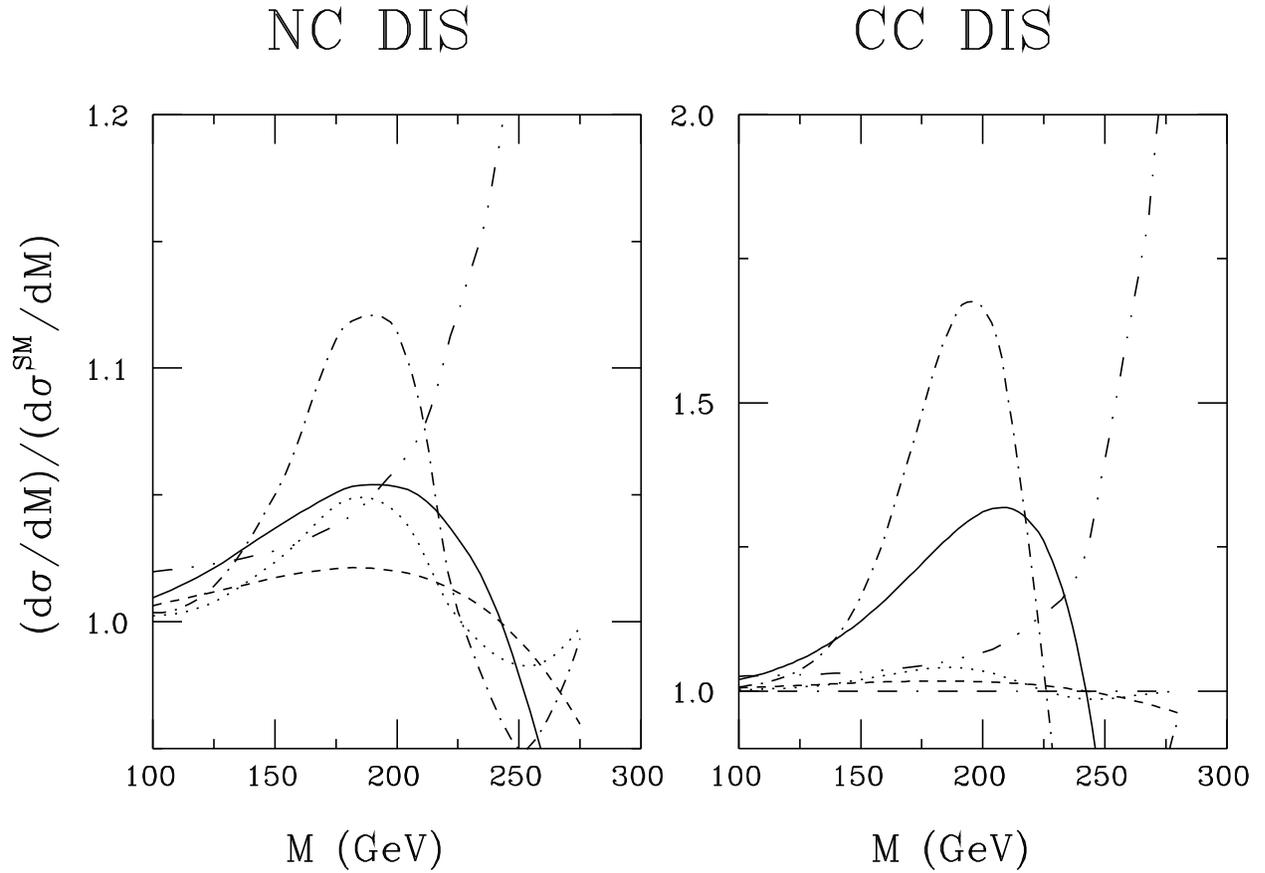}}
\bigskip
\caption{Predictions for $[d\sigma/dM]/[d\sigma^{SM}/dM]$
at HERA center-of-mass energy 
after integrating over $ y\geq 0.4$. Notation as for Fig.~\ref{rvsqsq};
the dash-dash-dot curve in the NC window is not
shown since it is nearly identical to the dash-dot-dot curve.}
\label{rvsm}
\end{figure}

In order to compare to HERA data, we have chosen to compute
the ratio of intrinsic charm plus the standard model perturbative prediction
from MRS96(R2) to the MRS96(R2) prediction alone 
(a) as a function of $Q^2$ after 
integrating over $0.1\leq y \leq 0.9$ and (b) as a function of $M$ after
integrating over $y\geq 0.4$, where $M=\sqrt{xs}$ is the $eq$
leptoquark mass. Experimental results
for these ratios have been presented by H1 \cite{h1}, and closely
related results have been presented by Zeus \cite{zeus}.
These ratios appear in Figs.~\ref{rvsqsq} and \ref{rvsm}, respectively.
Results for the $n=2$ and $n=8$ models are presented for $e^+p$ and $e^-p$,
NC and CC reactions. 

The $Q^2$ dependence in Fig.~\ref{rvsqsq} shows that
the $n=2$ and $n=8$ IC models predict $e^+p$ NC
enhancements that grow to roughly 6\% and 12\%, respectively, at high $Q^2$.
Much larger enhancements are indicated in the present HERA data.
If these large enhancements persist, they would have to have a 
primary source other
than intrinsic charm.  Further, the relatively small intrinsic charm
component could probably never be isolated from the dominant source
without charm tagging.
In the alternative scenario, where the enhancements decline
with increased statistics, it is clear that the
sensitivity to intrinsic charm could ultimately become very substantial.

Data from other reactions potentially provide further opportunity
for probing intrinsic charm. The additional possibilities include
$e^-p$ NC scattering and $e^{\pm}p$ CC scattering. The intrinsic
charm contribution to both NC and CC reactions is the same for
$e^+p$ and $e^-p$ scattering. The relative
IC enhancement in $e^-p$ NC scattering is then much smaller 
than in $e^+p$ NC scattering due to the increased
size of the SM cross section (deriving from the change in sign of the $F_3$
contribution). In contrast,
relative enhancements in $e^+p$ CC deep-inelastic
scattering will be much larger, exceeding 60\% and 150\% at
$Q^2>30000\gev^2$ for the $n=2$ and $n=8$ IC models, respectively. 
These larger enhancements occur because 
the SM $e^+p$ CC cross section is mainly proportional to the down quark
distribution and is, therefore, suppressed compared to the SM $e^+p$ NC
cross section, which is largely proportional to the much larger (at large $x$)
up quark distribution. 
Very small enhancements are predicted in $e^-p$ CC scattering.
(That the above systematics should apply has been
pointed out in Ref.~\cite{babu}.)

The (SM+IC)/SM ratios as a function of the leptoquark mass,
$M$, are particularly revealing. As seen in Fig.~\ref{rvsm},
the peak at moderate $x$ values in the intrinsic charm distribution
function (see Fig.~\ref{f2c}) results in a substantial 
peak in the (SM+IC)/SM ratio for $e^+p$ NC scattering
in the vicinity of $M\sim
200\gev$, \ie\ the same general location as the peak observed for data/SM
by H1.  Of course, the IC prediction
for the height of the peak 
[(SM+IC)/SM = 1.06 and 1.12 for the $n=2$ and $n=8$ IC models,
respectively]
is much smaller than that observed;
${\rm data}/{\rm SM} \sim 8$ at $M\sim 200\gev$ in the H1 data.
Increased statistics will decisively determine whether or not the
observed peak at $M\sim 200\gev$ in data/SM has 
anything to do with intrinsic charm. Observations of other reactions
could prove very valuable.
For example, the peaking in (SM+IC)/SM as a function of 
$M$ predicted by the IC models will be larger
in $e^+p$ CC scattering than in $e^+p$ NC scattering;
in the former case, the (SM+IC)/SM ratio reaches a maximum of 1.3 (1.7) in the
$n=2$ ($n=8$) IC models for $M\sim 200\gev$.

As a final point of comparison, we have also evaluated the effect of
adding an extra component, $\delta u(x)=0.02(1-x)^{0.1}$, to the valence
$u$ quark distribution function, as considered in Ref.~\cite{tung}.
Results for such an addition (after evolution) are also shown in
Figs.~\ref{rvsqsq} and \ref{rvsm}.  We observe that, in
the absence of other new physics contributions, data/SM as a function
of $M$ will allow one to easily discriminate between this possibility and 
intrinsic charm.

In conclusion, we find that intrinsic charm is unable to explain 
enhancements in the $e^+p$ neutral current cross section as large
as those seen at HERA, although a small peak in the vicinity of leptoquark
mass $M\sim 200\gev$ is a natural prediction. In the absence of
other new physics, HERA data will ultimately provide a sharp test
of the intrinsic charm picture, especially once
the charm component of $F_2$ is directly isolated.

\clearpage

\vspace{2cm}
\centerline{\bf Acknowledgments}
\vspace{.5cm}
This work was supported in part by the Director,
Office of Energy Research, Division of Nuclear
Physics of the Office of High Energy and Nuclear Physics
of the U.S. Department of Energy under
contracts No. DE-FG03-91ER40674 and DE-AC03-76SF0098, and
by the Davis Institute for High Energy Physics. We are grateful for
valuable conversations with S.J. Brodsky, B.W. Harris, J. Kiskis and W. Ko.

\end{document}